\addtocontents{}{}
\documentclass[11pt]{article}
\usepackage{amsmath,amssymb,color,graphics,epsfig,cite,footnote}
\usepackage{subfigure}
\usepackage{graphicx}
\usepackage{amsfonts}

\newcommand{\be}{\begin{equation}}
\newcommand{\ee}{\end{equation}}
\newcommand{\bea}{\setlength\arraycolsep{2pt} \begin{eqnarray}}
\newcommand{\eea}{\end{eqnarray}}
\newcommand{\nn}{\nonumber}

\def\ft#1#2{{\textstyle{\frac{\scriptstyle #1}{\scriptstyle #2} } }}

\def\0{{\sst{(0)}}}
\def\1{{\sst{(1)}}}
\def\2{{\sst{(2)}}}
\def\3{{\sst{(3)}}}
\def\4{{\sst{(4)}}}
\def\5{{\sst{(5)}}}
\def\6{{\sst{(6)}}}
\def\7{{\sst{(7)}}}
\def\8{{\sst{(8)}}}
\def\sst#1{{\scriptscriptstyle #1}}

\begin{document}

	
	\begin{center}
		{\Large {\bf Echoes from Asymmetric Wormholes and Black Bounce}}

		\vspace{20pt}
		
	{Min-Yan Ou, Meng-Yun Lai and Hyat Huang}

\vspace{10pt}

{\it College of Physics and Communication Electronics, \\
	Jiangxi Normal University, Nanchang 330022, China}

		\vspace{40pt}
		
		\underline{ABSTRACT}
	\end{center}

The time evolution of the field perturbations in the wormhole and black bounce backgrounds are investigated. We find that the asymmetry of spacetime results in the asymmetry of the effective potential of the perturbed equation. The quasinormal modes are strongly dependent on the shapes of the effective potentials. Specifically, the signals of echoes arise in some wormhole cases and reflect the asymmetric properties of wormholes. We examine the features of echoes within different circumstances. Besides, the negative values of effective potentials usually imply the instability of the system. By analyzing some specific metrics, we find that the negative regions of effective potentials are enclosed by the black hole horizons in these cases. But this statement could be broken in asymmetric cases.

\vfill {\footnotesize hyat@mail.bnu.edu.cn\ \ \ mengyunlai@jxnu.edu.cn\ \ \ \

	\thispagestyle{empty}
	
	\pagebreak
	\tableofcontents
\addtocontents{toc}{\protect\setcounter{tocdepth}{2}}
\newpage
	


\section{Introduction}\label{se1}

Black holes and wormholes are two typical fascinating solutions of General Relativity. Recently, progresses in astrophysical experiments have triggered a large interest in black hole physics. Two important progresses in this field are getting the black hole captured images by the Event Horizon Telescope  \cite{EventHorizonTelescope:2019dse,EventHorizonTelescope:2019uob,EventHorizonTelescope:2019jan} and detecting the gravitational waves (GWs) generated by the merger of black holes\cite{LIGOScientific:2016aoc}. Especially, there is a growing detected cases of GWs reported by LIGO Scientific and Virgo Collaboration\cite{LIGOScientific:2020iuh,LIGOScientific:2020zkf,LIGOScientific:2020stg}, which provides a lot of 
concrete examples to study the properties of black holes.

However, many studies point out that the probe of GWs ringdown signals has no sufficient evidence to claim that the signals come from black holes\cite{Abramowicz:2002vt,Cardoso:2016rao}. In Ref.\cite{Cardoso:2016rao}, it was suggested that the horizonless objects can mimic black holes in the sense of initial ringdown signals. The compact horizonless objects involve neutron stars\cite{Cardoso:2019rvt}, boson stars\cite{Schunck:2003kk} and wormholes\cite{Cardoso:2016rao,Bueno:2017hyj,Liu:2020qia} and so on. Due to the widespread interests of wormhole physics\cite{Lobo:2005us,Almheiri:2019qdq,Maldacena:2017axo,Nandi:2006ds,Kanti:2011jz,Blazquez-Salcedo:2020czn,Huang:2020qmn}, it has a great motivation to distinguish black holes with wormholes. A useful way to achieve this goal is perturbing the objects and observing their evolution.
When black holes or wormholes are perturbed, they radiate GWs based on the initial conditions of the perturbations. After that, they reduce to damped oscillations with complex frequencies. The modes of oscillations are the so-called quasinormal modes (QNMs), which encode information on the evolutions. Researches on the QNMs of various black hole and wormhole solutions are quite prominent\cite{Berti:2009kk,Cardoso:2003vt,Cardoso:2017soq,Konoplya:2016hmd,Konoplya:2018ala,Aneesh:2018hlp,Blazquez-Salcedo:2018ipc}. There is an intriguing phenomenon called \textit{echoes} in QNMs. The echoes are generated by the reflected waves of the system, in which a sudden increase of amplitudes in QNM spectrums appears. We should note, however, that not all black holes or wormholes can generate echoes after perturbations. An interesting direction is to use the features of echoes to examine the properties of black holes and wormholes. For example, Ref.\cite{DAmico:2019dnn,Liu:2021aqh} pointed out that echoes arise from black holes if there are local Lorentz symmetry violations or discontinuity in perturbation effective potentials. Ref.\cite{Liu:2020qia} studied the echoes of two wormhole models, they found there are obvious echo signals in wormhole geometries.

On the other hand, investigating the differences between black holes with wormholes in the same circumstance is necessary. To do this, we need to find the theories that admit both the black hole and wormhole solutions. A natural consideration is regular black holes. A regular black hole means there is no singularity inside the event horizons. Simpson and Visser proposed a new concept named black bounce recently\cite{Simpson:2018tsi}. This new concept was inspired by Bardeen’s regular black hole and similar to Bronnikov’s dark universe\cite{Bronnikov:2006fu}. Simpson and Visser suggested a two parameters metric, which can describe traversable wormhole and black bounce regarding the values of the parameters. The black bounce mechanism and the Simpson-Visser (SV) metric have caused much more concern in physicists. Later, Huang and Yang (HY) obtained a charged wormhole/black bounce solution in the Einstein-Maxwell-scalar theory\cite{Huang:2019arj}. Then Lobo \textit{et al.} constructed a novel black bounce spacetimes (LRSSV metric)\cite{Lobo:2020ffi}. The observational properties, such as the gravitational lensing\cite{Islam:2021ful,Cheng:2021hoc,Tsukamoto:2020bjm} and perturbation echoes\cite{Churilova:2019cyt,Yang:2021cvh}, of SV metric and the LRSSV metric are studied. Especially in Ref.\cite{Churilova:2019cyt}, they investigated the SV metric and suggested that the black hole/wormhole transition is characterized by the echoes. A similar conclusion is presented in the LRSSV metric\cite{Yang:2021cvh}. 

It should be mentioned that the SV metric and the LRSSV metric are both symmetric. It means the two parts of the spacetime regions are double copies. Then the symmetry of the metrics causes the symmetric effective potentials of the perturbations. Being distinct from the above symmetric metrics, HY metric involve asymmetric cases. It is worth examining that how an asymmetric metric does after the perturbations. 

In this paper, we would like to draw attention to the evolution of (a)symmetric black bounce solutions when they are perturbed. We found in symmetric black bounce metrics (SV metric and LRSSV metric and a special case in HY metric), the negative regions of effective potentials are enclosed by black hole horizons. But move to the HY metric, the asymmetry of metrics raises asymmetric effective potentials, such that the negative regions could be outside of the black hole horizons. Furthermore, focusing on the effective potentials, the asymmetry makes a significance of the QNM spectrums. In Ref.\cite{Churilova:2019cyt,Yang:2021cvh}, they list all the possible shapes of the effective potential in the SV metric and the LRSSV metric. We choose two common classical shapes of all the possible effective potentials, namely the single peak and the two peak forms, to compare the difference. Note that the single peak form does not produce any echoes, we mainly care about the two peaks form. For the symmetric metrics, we can only adjust the deep of the well between the two peaks. Nevertheless, it can change the related high of the two peaks in the asymmetric metrics. The results lead us to a deeper understanding of time evolution for the perturbations. 

The paper is organized as follows. In Sec.\ref{bri}, we make brief reviews of the symmetric and the asymmetric wormhole/black bounce metrics. In Sec.\ref{se2}, we deal with the perturbations and derive the perturbation equations for general spherically symmetric metrics. Then in Sec.\ref{se3}, we present the finite difference method to solve those perturbation equations. The effective potentials with different metrics are analyzed in Sec.\ref{se4}. The results of time evolution for the perturbations are presented in Sec.\ref{se5}.
Finally, we conclude our paper in Sec.\ref{con}.

\section{Review of black bounce solutions}\label{bri}

A spherically symmetric ansatz with wormhole throat-like geometry takes the form
\be\label{ansazt}
ds^2=-h(r)dt^2+h^{-1}(r)dr^2+(r^2+q^2)d\Omega^2_{2},
\ee
where
\be
d\Omega^2_{2}=d\theta^2+\sin^2 \theta d\phi^2.
\ee
For the Simpson-Visser (SV) metric, we have\cite{Simpson:2018tsi}
\be\label{viss}
h(r)=h_{sv}=1-\ft{2m}{\sqrt{r^2+q^2}}.
\ee
This metric describes traversable wormhole or black bounce up to the value of parameter $q$. 
which can be used to describe traversable wormhole for $q>2m$ or regular black hole for $0<q<2m$. It is obvious the metric has $r$ to $-r$ symmetry, such that it depicts a symmetric wormhole or regular black hole. In the regular black hole case, there is a wormhole throat-like geometry in $r=0$: the wormhole throat usually is a timelike hypersurface while it is spacelike. Simpson and Visser treat this geometry as a bounce into a future incarnation of Universe\cite{Simpson:2018tsi}.

There is another symmetric metric proposed by Lobo \textit{et al.}\cite{Lobo:2020ffi}. They consider a charged solution from the Einstein gravity with the matter sector as an anisotropic fluid. The LRSSV metric is given by the same form with \eqref{ansazt} and 
\be\label{lobo}
h(r)=h_{l}=1-\ft{2m}{(r^4+q^4)^{1/4}}.
\ee
The properties of this metric are very similar to the SV metric. It has the $r$ to $-r$ symmetry too and describes symmetric spacetimes. When $q>2m$, it depicts charged a traversable wormhole; When $0<q<2m$, it depicts a black bounce.

Huang and Yang (HY) introduced a phantom scalar as the matter sector in the Einstein-Maxwell-phantom scalar theory \cite{Huang:2019arj}, then obtained an asymmetric black bounce solution. The Lagrangian for the theory is given by 
\bea\label{lag}
&&\mathcal{L}=\sqrt{-g}(R+\ft{1}{2}(\partial \phi)^2-\ft{1}{4}Z^{-1}F^2).\nn\\
&&Z=\gamma_1 \cos \phi+\gamma_2 \sin\phi.
\eea
The phantom field is defined by flapping the sign of kinetic term in Lagrangian, then it is easy to see the scalar field is  phantom-like everywhere. When the coupling function $Z$ becomes negative somewhere, the Maxwell field is also phantom-like. The phantom fields as exotic matters that support wormhole throats.

Within the ansatz \eqref{ansazt}, the HY solution is given by
\bea\label{ax}
&& h(r)=1-\ft{\gamma_2 Q^2 r}{4q(r^2+q^2)}+\ft{\gamma_1 Q^2}{4 (r^2+q^2)},\qquad \phi=\phi(r,q), \quad A=\xi(r,q,Q)dt,\nn\\
&&\phi=2\arccos(\ft{r}{\sqrt{r^2+q^2}}),\qquad \xi'=\ft{QZ}{r^2+q^2},
\eea
 where $r\in (-\infty,+\infty)$. It follows that the solution\eqref{ax} is asymptotic to flat in $r\to \pm\infty$.  
It is easy to  reduce the solution to the Ellis wormhole solution in the limit $Q\to0$. We need to emphasis that we won't take the parameter $q$ to be zero in this paper. The mass $M$ and the electric charge $Q_e$ of the solution are \cite{Huang:2019arj}
\be
M=\ft{\gamma_2Q^2}{8q},\qquad Q_e=\gamma_1 Q.
\ee
It characterize traversable  wormholes, if and only if 
\be\label{conwh}
Q^2<\ft{8q^2(\gamma_1+\sqrt{\gamma_1^2+\gamma_2^2})}{\gamma_2^2}.
\ee
Although the two asymptotic regions at $r\to\pm\infty$ are both flat, the spacetime curvatures are different in the vicinity of the wormhole throat. \textit{It means that the HY metric is an asymmetric solution. }
 
When
\be\label{bhcon}
 Q^2\geq\ft{8q^2(\gamma_1+\sqrt{\gamma_1^2+\gamma_2^2})}{\gamma_2^2},
\ee
the HY metric  characterize regular black holes. In fact, this situation can depict two types of regular black holes. Type I regular black hole looks like an RN black hole, which has outer and inner horizon but the curvature singularity at $r=0$ is replaced by a wormhole throat.  Type II regular black hole is black bounce where the bounce occurs at $r=0$ and the two horizons are located on both sides of the bounce. 

Specially, if $\gamma_2=0$, the function $h$ in metric \eqref{ax} reduces to 
\be
h=1-\ft{ Q^2}{4 (r^2+q^2)}
\ee
where we set $\gamma_1=-1$ without loss of generality. This is analogy with SV metric \eqref{viss} and the LRSSV metric \eqref{lobo}. When $2q<Q$, it depicts a black bounce; when $2q=Q$, it depicts a one-way wormhole; when $2q>Q$, it depicts a traversable wormhole.

\section{Perturbations of black holes}\label{se2}

In general, there are usually three types of perturbations, namely scalar, electromagnetic and gravitational field perturbations. One of the goals of this work is to discuss how the asymmetry of spacetimes affects the results of the perturbations. Note that the gravitational fields and the scalar fields behave qualitatively with each other. And hence we study the scalar field perturbations as a proxy for the gravitational perturbations\cite{Liu:2020qia}. 

The equations of motion of a massless scalar field $\phi$ is given by
\be\label{eoms}
\ft{1}{\sqrt{-g}}\partial_\mu(\sqrt{-g}g^{\mu\nu}\partial_\nu\phi)=0.
\ee
Using the way of separation of variables, we could separate the scalar field $\phi$ into two part, which are the spherical harmonics $Y_{l,m}$ and the radial part $\psi (t,r)$. 
\be
\phi=\underset{l,m}{\Sigma}\ft{\psi(t,r)}{\sqrt{r^2+q^2}}Y_{l,m},
\ee
where $l$ denote angular number and $m$ denote azimuthal number. Then the E.O.M of scalar field \eqref{eoms} reduces to
\bea
 &&-\frac{\partial^2 \psi(t,r)}{\partial t^2} + \frac{1}{h(r)^2} \frac{\partial^2 \psi(t,r)}{\partial r^2} + h(r) h^{\prime}(r) \frac{\partial \psi(t,r)}{\partial r}\nn\\
 &&-\frac{h(r) \bigg( q^2 h(r)+(q^2+r^2)(l(l+1)+r h^{\prime}(r))\bigg)}{(q^2+r^2)^2} \psi(t,r)=0.
 \eea
For the sake of research, we use the tortoise coordinate $r_*$ replace $r$. The tortoise coordinate $r_*$ here is defined by
\be
dr_*=\ft{1}{h(r)}dr.
\ee
It is also worth mentioning that the tortoise coordinate $r_*$ is monotonically increasing with $r$ but avoids the horizon and more conveniently to examine the QNMs.
And hence (11) becomes wave function in the time domain, namely
\be\label{pur}
-\ft{\partial^2 \psi(t,r)}{\partial t^2}+\ft{\partial^2 \psi(t,r)}{\partial r_*^2}-V(r) \psi(t,r)=0,
\ee
where
\begin{equation}
    V(r)=\frac{h(r)\bigg(q^2 h(r)+(q^2+r^2)(l(l+1)+r h^{\prime}(r))\bigg)}{(q^2+r^2)^2}.
\end{equation}
Furthermore, we could rewrite $\psi(t,r)$ as $\Psi(r) \exp(-i \omega_{l,n}t)$ ,where $e^{-i \omega t}$ is the time evolution of the scalar field. Then we obtain wave function in the frequency domain, namely
\be
\ft{d^2\Psi(t,r)}{dr_*^2}+(\omega_{l,n}^2-V(r))\Psi(r)=0.
\ee
where $n$ denotes the overtone number. The  $\omega_{l,n}$ is a complex number characterize the frequency of QNMs. 

The effective potential, as know as the Regge-Wheeler potential, can rewrite to the following form
\be\label{effectiveP}
V(r)=h(r)F(r),
\ee
with
\be \label{Fr}
F(r)=\ft{1}{(q^2+r^2)^2}\bigg( q^2 h+(q^2+r^2)(l+l^2+r h'(r))\bigg).
\ee

As we will see as follows, the time evolution of the perturbed fields is strongly related to the effective potential $V$. 

\section{Finite difference method}\label{se3}

There are two common methods to solve the perturbation equation \eqref{pur} numerically, namely WKB approximations and the Finite difference method. In this work, we follow the last one, it also named time-domain integration.
The method we used here could be found in Ref.\cite{Liu:2020qia}

We discretize the coordinates $t=i \Delta t$ and $r_*=j \Delta r_*$, where the i and j are integers. Then $\psi(t,r_*)=\psi(i\Delta t,j\Delta r_*)$, we make $\Delta t$ and $\Delta r_*$ to the constant. Then we can replace $\psi(i\Delta t,j\Delta r_*)$ with $\psi_{i,j}$. In a similar way, the $V(r_*)= V(j \Delta r_*)= V_j$. Then the 
discretized form of equation \eqref{pur} becomes
\bea
&&-\frac{\psi(i+1,j)-2\psi(i,j)+\psi(i-1,j)}{\Delta t^2}
+\frac{\psi(i,j+1)-2\psi(i,j)+\psi(i,j-1)}{\Delta r_*^2}-V(j)\psi(i,j)=0.
\eea
As same as Ref.\cite{Zhu:2014sya}, we considering the initial Gaussian distribution $\psi(t=0,r_*)=e^{- \frac{(r_*- \bar{a})^2}{2 b^2}}$ and $\psi(t<0,r_*)=0$. In the present work, we take $b=1$ and the value of $\bar{a}$ will be chosen accordingly. Then we can derive the evolution of $\psi$ by
\bea\label{dis}
\psi(i+1,j)=-\psi(i-1,j)+\bigg(2-2 \frac{\Delta t^2}{\Delta r_* ^2}- \Delta t^2 V(j) \bigg) \psi(i,j) 
+ \frac{\Delta t^2}{\Delta r_* ^2}  \bigg(\psi(i,j+1)+\psi(i,j-1) \bigg).
\eea
In this work, we set the $\Delta t / \Delta r_*=0.5$ to ensure that the von Neumann stability
condition can be fullfilled\\

\section{Effective potential }\label{se4}

Before we solve the discretized equation \eqref{dis} with numerical methods above and show the main results in the next section, it is necessary to discuss the effective potentials $V$ with different metrics.
  
Based on the \eqref{effectiveP}, the behaviors of the effective potential $V(r)$ are determined by the roots of the following three equations:
\bea
h(r)&=&0, \label{hEq}\\
F(r)&=&0, \label{FEq}\\
\frac{d V(r)}{dr}&=&0, \label{DVEq}
\eea
where the roots of Eq.~\eqref{hEq} determine the existence and the position of the horizons, the roots of Eq.~\eqref{hEq} and Eq.~\eqref{FEq} describe the transformations of the sign of $V(r)$, and the roots of Eq.~\eqref{DVEq} represent the extreme points of $V(r)$. 

When effective potential $V$ becomes negative at somewhere, as the study in Ref.\cite{Myung:2018jvi}, the growing QNMs could appear. This is a signal of the instability of the systems. In what follows, we will point out that there are no negative regions outside the black hole horizons in the cases of symmetric black bounce. But this argument could be broken in asymmetric black bounce cases.

\subsection{Symmetric spacetimes}

As we mentioned above, the SV metric, the LRSSV metric, and the HY metric with $\gamma_2=0$ are both symmetric solutions. It is worth noting that $\gamma_2=0$ corresponds to the solution with zero mass. A symmetric wormhole with zero mass is acceptable, such as the famous Ellis wormhole\cite{Ellis:1973yv}. 

As the conclusions of Ref.\cite{Churilova:2019cyt}, there are two shapes of effective potentials in the SV metric. The two peaks shapes happen only in traversabale certain wormhole cases, namely $q>2m$. It follows that the black bounce cases have only single peak shapes of the effective potential and hence there are no echoes. The LRSSV metric holds a similar conclusion, besides it arises a new three peaks shape in the traversable cases. Let's expand the effective potential $V$ at the two asymptotic regions, namely $r\to\pm\infty$, we found the above two symmetric metrics have the same form. For any $l\neq0$ waves perturb the spacetime, they give rise to the same leading term of the effective potential $V$, namely
\be\label{Vl}
V(\pm\infty)=\ft{l(l+1)}{r^2}+\cdots
\ee
When $l=0$, the leading term changes to
\be
V(\pm\infty)=2m(\ft{1}{r^2})^{3/2}+\cdots
\ee
Keeping this in mind, it is clear that the effective potentials are trend to $0^+$ no matter in $r\to\pm\infty$. We would use this conclusion to prove that there are no negative regions of the effective potentials outside of the horizons in the cases of the SV metric, the LRSSV metric, and the HY metric of $\gamma_2=0,\gamma_1<0$.

The SV metric, the LRSSV metric and the HY metric of $\gamma_2=0,\gamma_1<0$ share the following form of $h(r)$:
\be
h(r) = 1- \frac{A}{(r^{2N}+q^{2N})^B}
\ee
where $A>0$, $B>0$ and $N$ is a positive integer. Let us assume that the roots of $h(r)$ locate at $r=\pm r_0$. Then we have $h(r)>0$ for $r>|r_0|$ and $h(r)<0$ for $r<|r_0|$, since $h(r)>0$ for $r\rightarrow \pm \infty$. On the other hand, it can be easily verified that $rh'(r) \geq 0$. Therefore the roots of $F(r)=0$ can only appear in the region $r\leq |r_0|$, and $F(r_0)=0$ only happens for $r_0 =0$ and $l=0$.

From the above arguments, we arrive at the conclusion that the transformation of the sign of $V(r)$ only happens between the two horizons. Combining with the asymptotic behavior of $V(r)$, it is positive when $r\rightarrow \pm\infty$. Because for $l\neq0$, the leading term of $V$ at the boundaries are the same with \eqref{Vl}. For $l=0$, the leading term becomes
\be\label{veff2}
V(\pm\infty)=\ft{q^2-Q^2 \gamma_1/2}{r^4}+....
\ee
Thus the negative effective potential only appears between the two horizons. 

To sum up, we show that there are no negative regions of the effective potentials outside the horizons in all the three symmetric metrics. We use the illustrations, varying with $q$, to exhibit our conclusion more intuitively in Fig.~\ref{roots}. For a certain value of $q$, the red dashed line denotes how many peaks and wells of the effective potential $V(r)$ while the blue solid line denotes how many horizons. Based on such profiles, we found for a large $q$, the three metrics give rise to traversable wormholes that possess a single peak of effective potential at the wormhole throat. When $q$ decreases, more peaks emerged. To be specific, there are two peaks in the SV metric and the HY metric with $\gamma_2=0$ and three peaks in the LRSSV metric. For a small $q$, there are two symmetric black hole horizons emerge and each side of the horizon has only one peak.
\begin{figure}[htbp]
\centering
\subfigure[$l=0$]{
\includegraphics[width=0.45\textwidth]{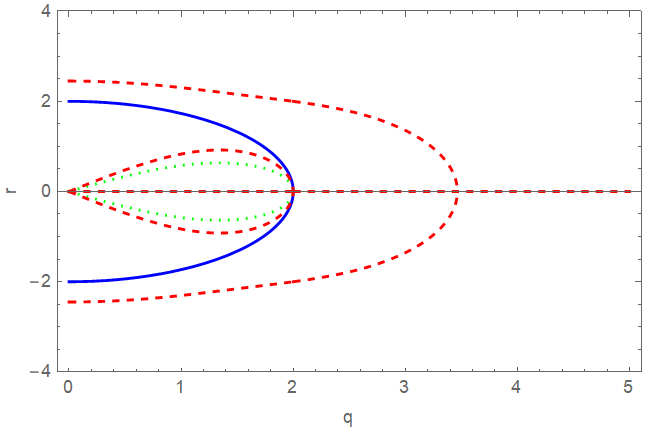}
}
\quad
\subfigure[$l=1$]{
\includegraphics[width=0.45\textwidth]{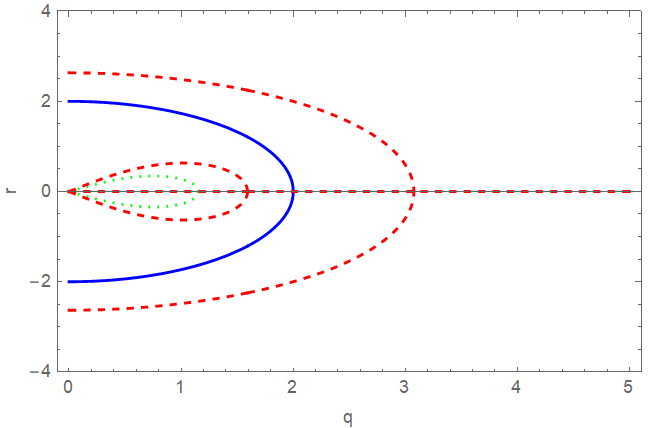}
}
\quad
\subfigure[$l=0$]{
\includegraphics[width=0.45\textwidth]{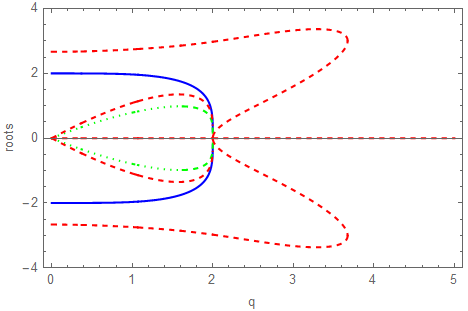}
}
\quad
\subfigure[$l=1$]{
\includegraphics[width=0.45\textwidth]{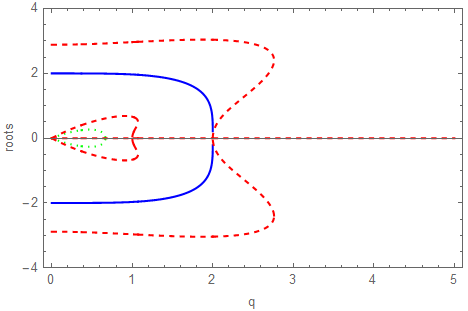}
}
\quad
\subfigure[$l=0$]{
\includegraphics[width=0.45\textwidth]{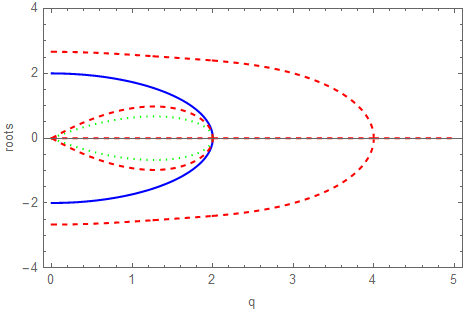}
}
\quad
\subfigure[$l=1$]{
\includegraphics[width=0.45\textwidth]{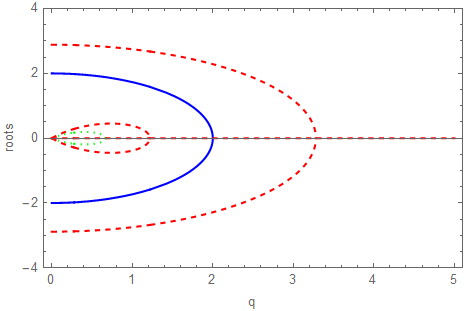}
}
\caption{\it Roots of \eqref{hEq} (blue solid line), \eqref{FEq} (green dotted line) and \eqref{DVEq} (red dashed line) of the three symmetric metrics. From top to bottom, they correspond to the SV metric, the LRSSV metric and the HY metric with $\gamma_2=0$ respectively. The left column shows the case of $l=0$ while the right column shows the case of $l=1$. }\label{roots}
\end{figure}

\subsection{Asymmetric spacetimes}

When $\gamma_2\neq0$, the HY metric describes an asymmetric spacetime. It should be noted that the asymptotic behavior in $r\to\pm\infty$ of $V(r)$ becomes
\be\label{veff3}
V(\pm\infty)=\ft{l(l+1)}{r^2}-\ft{(-1+l+l^2)Q^2 \gamma_2}{4q r^3}+\cdots
\ee
If $l\neq0$, the leading term is the same with the above symmetric cases and the effective potential $V(r)$ in two asymptotic regions are both trends to $0^+$. But in the cases of $l=0$,  we find
\be
V(\pm\infty)=\ft{Q^2 \gamma_2}{4q r^3}+\cdots
\ee
It follows that $V(+\infty)$ and $V(-\infty)$ have always different signs. In fact, $V(+\infty)$ trends to $0^+$ and $V(-\infty)$ trends to $0^{-}$. Compare with the symmetric cases in the above section, we may conclude that, \textit{for the s-wave ($l=0$) perturbations, the asymmetry of spacetime cause negative regions of the effective potentials lie in the outside horizons .}

In this work, we concentrate on the black bounce and wormhole cases and hence ignore the RN-like black hole cases. Moreover, we do not interest in the structures inside black hole horizons. Within these assumptions, there are only single peak shapes and two peaks shapes. These two shapes are good examples to illustrate the asymmetry that affects the time evolution of the perturbations. 
The single peak cases involve all the black bounce cases and some wormhole cases. We present the $V(r_*)$ with various $\gamma_2$ in Fig.~\ref{singlepeak}. It demonstrates that the increases of $\gamma_2$  aggravate the asymmetry. 

The two peaks cases arise in certain traversable wormhole geometries. The left column of Fig.~\ref{gamma2change}, \ref{lchange}, \ref{Mchange} and \ref{Qchange} show the shapes of these two peaks cases with different parameters. We should emphasize that the implications of each parameter are different. The $\gamma_2$ arises in the coupling function $Z(\phi)$ of the Lagrangian. The $l$ is the eigenvalue of the spherical harmonics $Y_{l,m}$, which corresponds to the angular momentum of the perturbation fields. The scalar parameter $q$ relates to the mass $M$ of the wormhole when we fix $Q$. The parameter $Q$ is proportional to the electrical charge $Q_e$. 

\section{Qusinormal modes}\label{se5}

We list the results of the time evolution profiles in single peak and two peaks cases in this section. For the single peak profile, as we present in Fig.~\ref{singlepeak}, we compare the different asymmetric cases with various $\gamma_2$. Although there are no echo signals, we find that the asymmetry affects the modes of oscillations and the final tendencies are stable.
\begin{figure}[htbp]
\centering
\subfigure[Effective potentials]{
\includegraphics[width=0.45\textwidth]{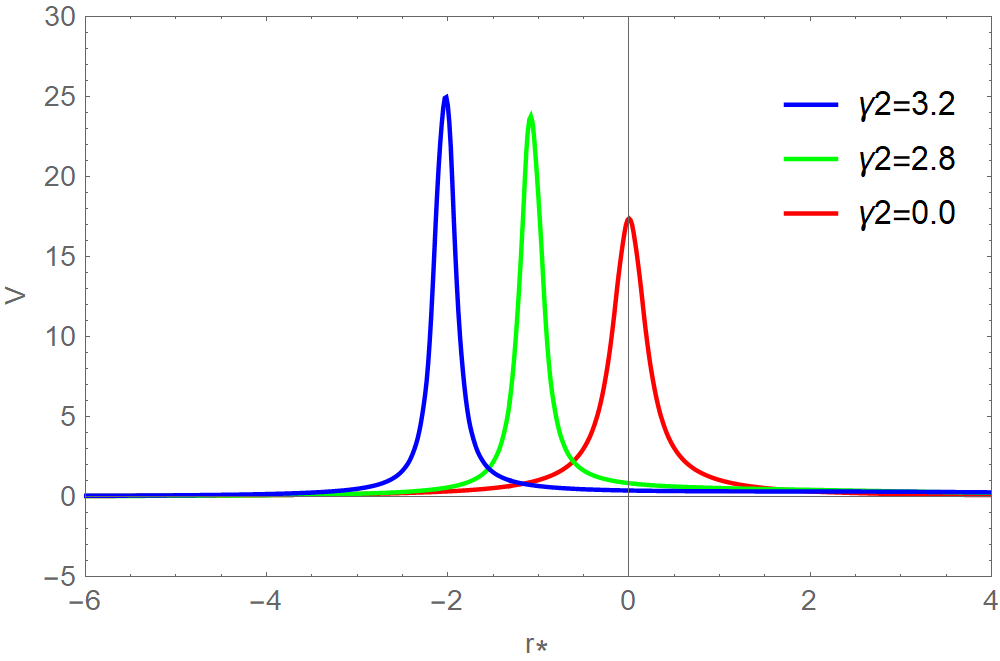}
}
\quad
\subfigure[Time evolution of perturbations]{
\includegraphics[width=0.45\textwidth]{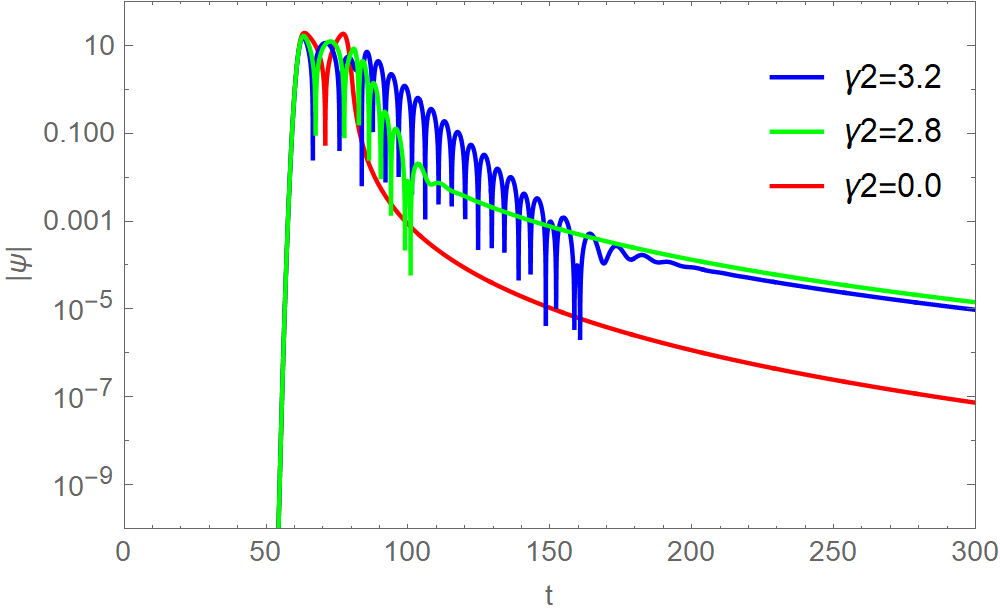}
}
\caption{\it The effective potentials and the corresponding time evolution profiles with different  $\gamma_2$. We set $q=0.6,\gamma_1=1,l=1,Q=1$.}\label{singlepeak}

\end{figure}

\begin{figure}[htbp]
\centering
\subfigure[Effective potentials]{
\includegraphics[width=0.45\textwidth]{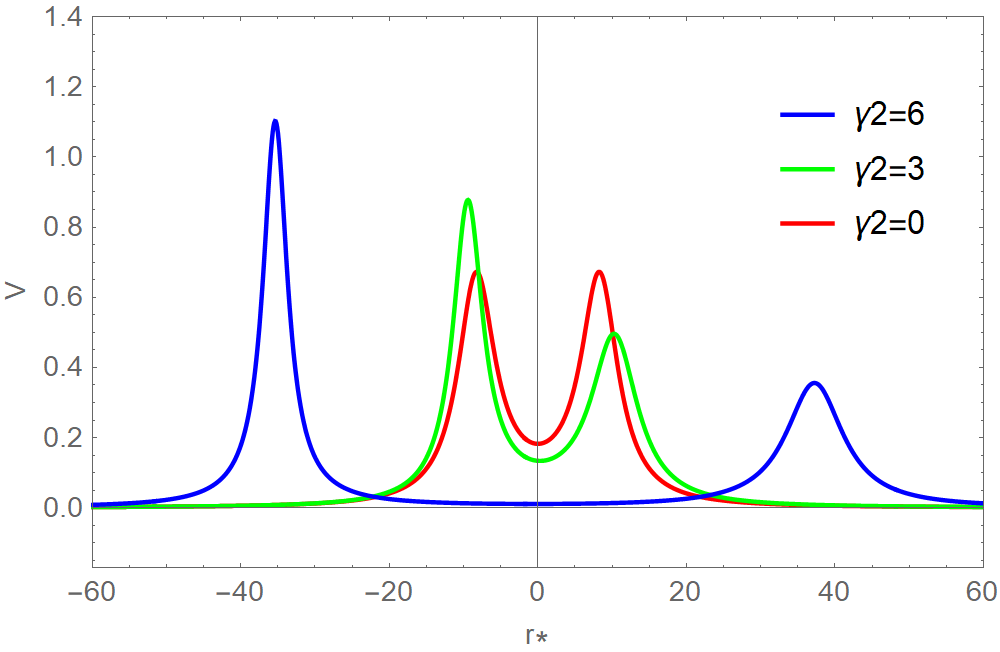}
}
\quad
\subfigure[Time evolution of perturbations]{
\includegraphics[width=0.45\textwidth]{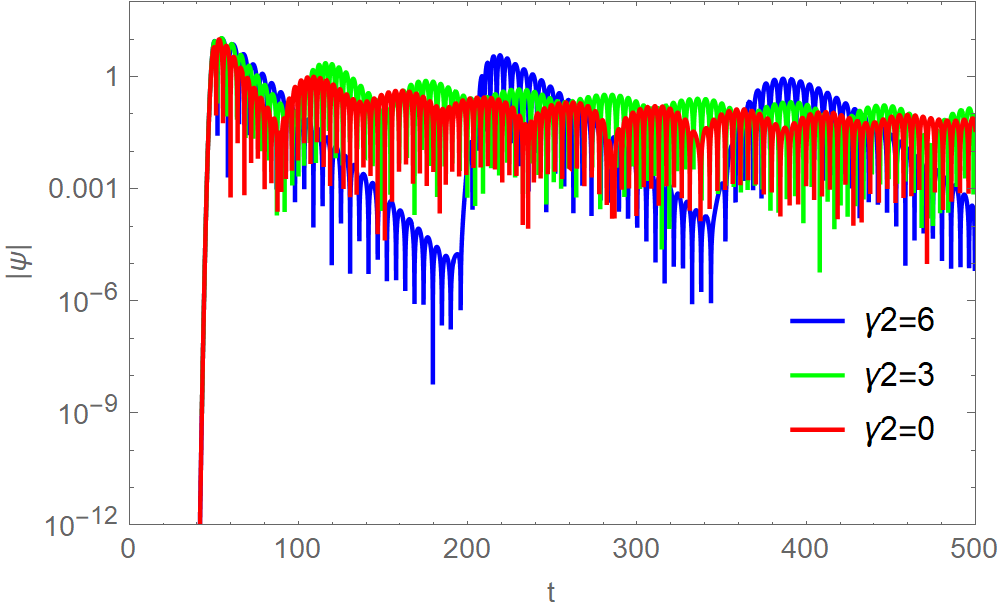}
}
\caption{\it  The effective potentials and the corresponding time evolution profiles with different $\gamma_2$. We set $q=1.65,\gamma_1=-10,l=2,Q=1$.}\label{gamma2change}
\end{figure}
\begin{figure}[htbp]
\centering
\subfigure[Effective potentials]{
\includegraphics[width=0.45\textwidth]{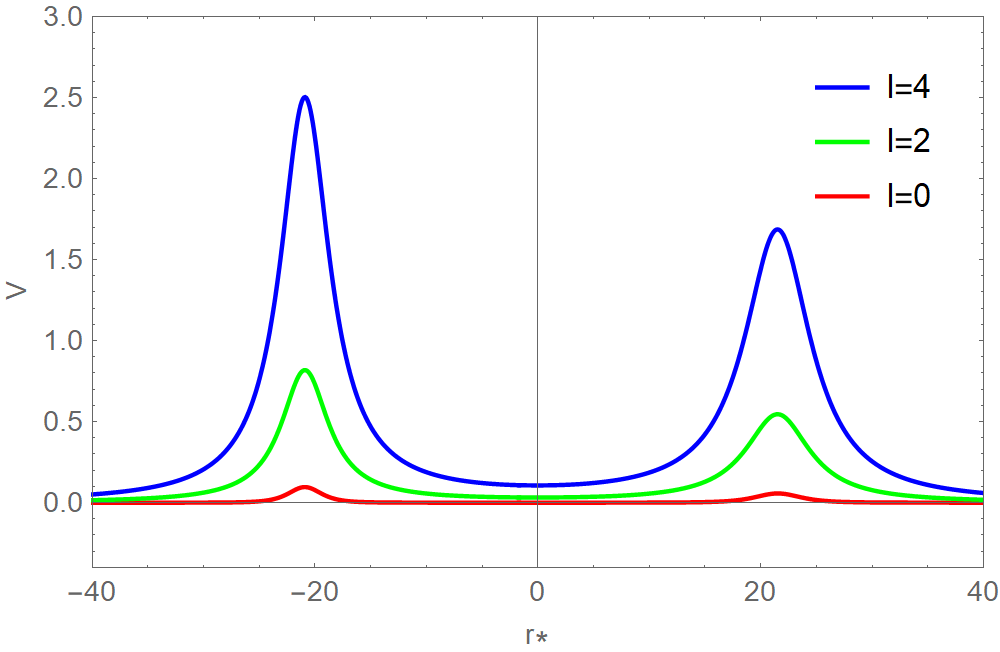}
}
\quad
\subfigure[Time evolution of perturbations]{
\includegraphics[width=0.45\textwidth]{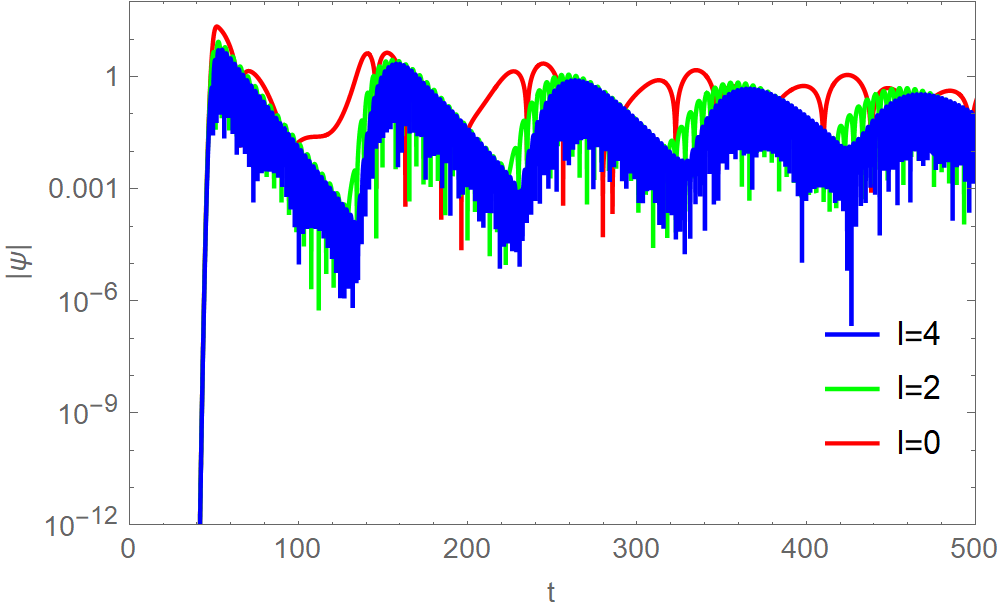}
}
\caption{\it  The effective potentials and the corresponding time evolution profiles with different $l$. We set $q=1.6,\gamma_1=-10,\gamma_2=2,Q=1$.}\label{lchange}
\end{figure}
\begin{figure}[htbp]
\centering
\subfigure[Effective potentials]{
\includegraphics[width=0.45\textwidth]{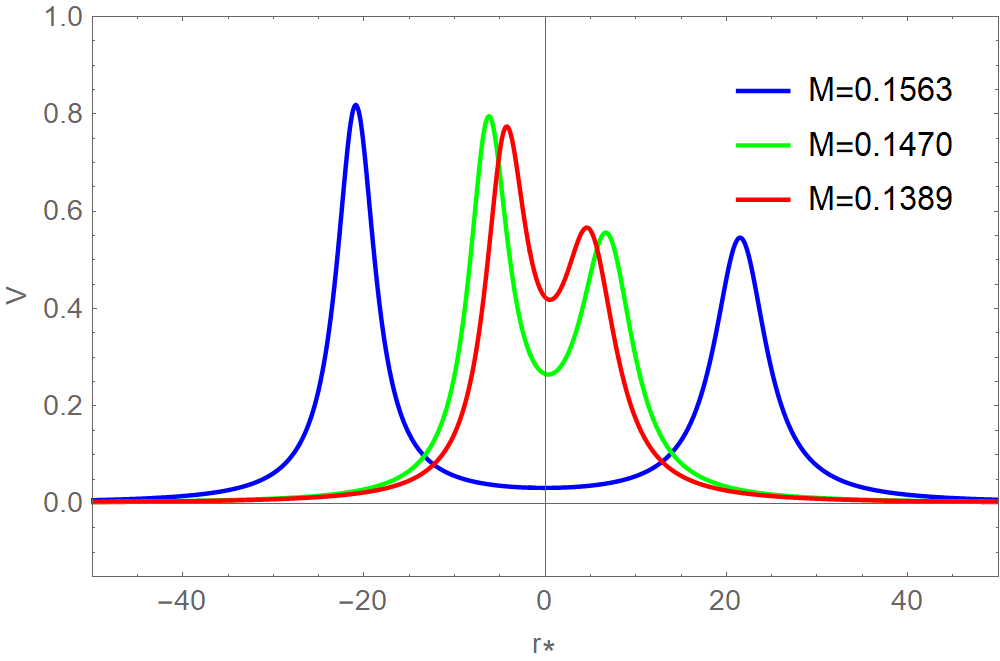}
}
\quad
\subfigure[Time evolution of perturbations]{
\includegraphics[width=0.45\textwidth]{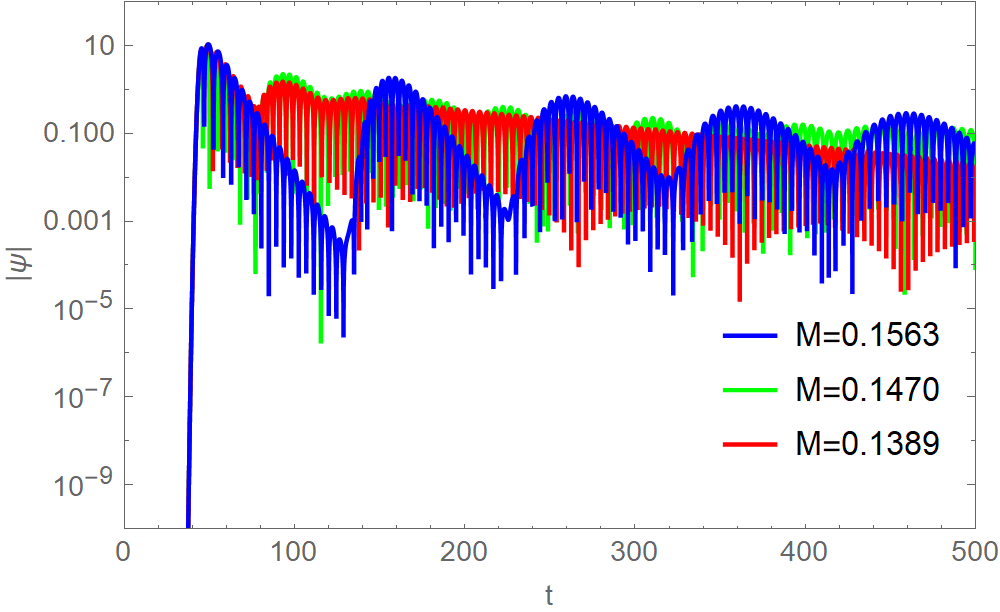}
}
\caption{\it  The effective potentials and the corresponding time evolution profiles with different $M$. We set $\gamma_1=-10,\gamma_2=2,l=2,Q=1$.}\label{Mchange}

\end{figure}
\begin{figure}[htbp]
\centering
\subfigure[Effective potentials]{
\includegraphics[width=0.45\textwidth]{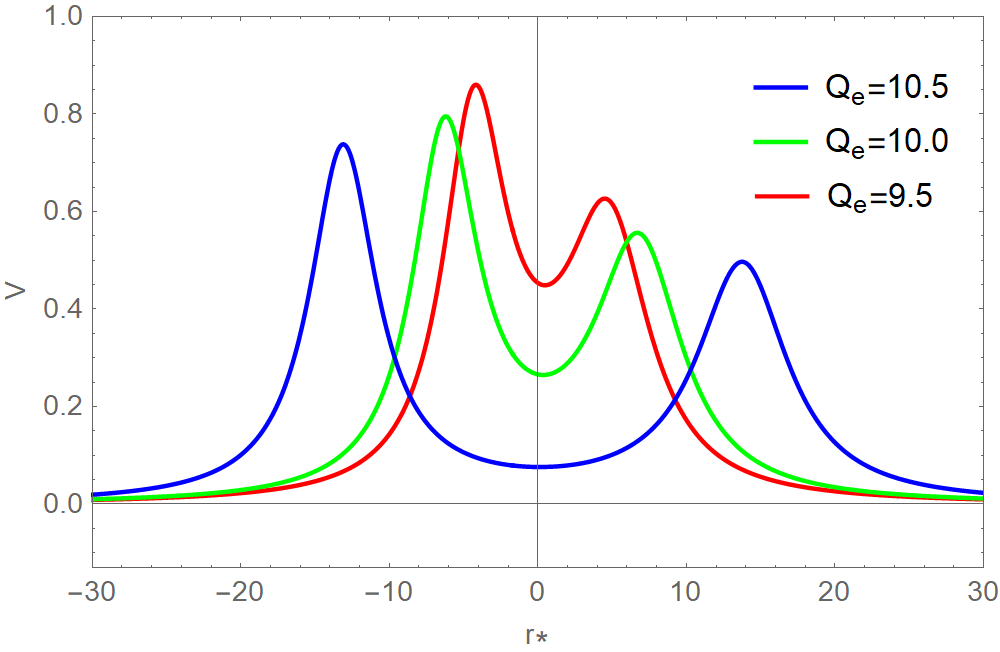}
}
\quad
\subfigure[Time evolution of perturbations]{
\includegraphics[width=0.45\textwidth]{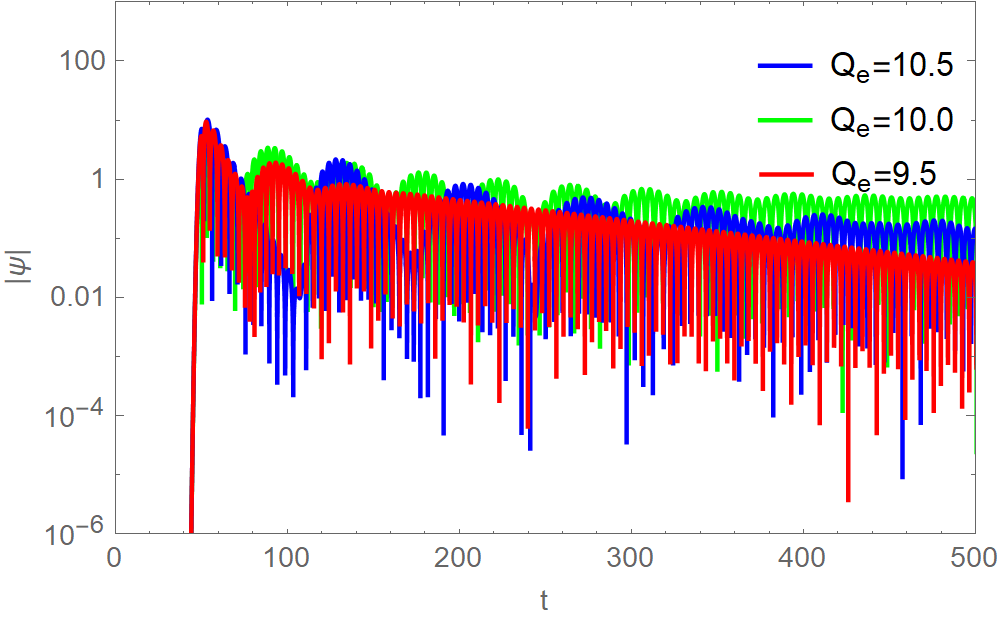}
}
\caption{\it  The effective potentials and the corresponding time evolution profiles with different $Q_e$. We set $q=1.7,\gamma_1=-10,\gamma_2=2,l=2$.}\label{Qchange}

\end{figure}

For the two peaks profiles, as we present in Fig.~\ref{gamma2change}, \ref{lchange}, \ref{Mchange} and \ref{Qchange}, we classify the situations with different $\gamma_2, l, M, Q_e$. Two peaks cases can produce echoes.
The left column shows the effective potential $V(r_*)$ while the right column is the corresponding time evolution profiles. Based on those profiles, we sum up our results as follows:

\begin{itemize}

    \item Theoretical parameter $\gamma_2$ relates to the coupling strength between the scalar field and the electromagnetic field. When $\gamma_2=0$, there are two symmetric peaks of $V$. With the increase of $\gamma_2$, these two peaks become further and further such leading to a broad well. We should note one of the peaks becomes higher while another peak becomes lower. The corresponding QNMs profile suggests that larger $\gamma_2$ has lower frequency but larger amplitude echoes. The first echo signal after the initial ringdown arises as sooner as smaller $\gamma_2$. One may expect to observe large amplitudes when $\gamma_2$ becomes large.
    
    \item When we use the scalar waves with different angular momentum $l$ to perturb the same wormholes, it gives rise to the changes of the high of the two peaks. Larger $l$ gives rise to higher peaks. Interestingly, changing with $l$ does not affect the width of the two peaks. According to the QNMs profile, we can see smaller $l$ has triggered higher frequencies of the echoes.  The amplitudes of the echoes become smaller with larger $l$. With the smaller $l$, the sooner detection of the first echo signal after the initial ringdown in this case.
    
    \item The wormhole mass $M$ strongly affects the width of the two peaks. An intriguing feature is that changing the mass of the wormholes does not significantly alter the high of the two peaks. A Larger mass leads to broader width and deeper well. The corresponding QNMs profile implies that larger mass wormholes have larger amplitude and lower frequency echoes. The wormhole mass relates to the parameters $Q$ and $q$. But we fixed $Q$ and hence varying $q$ to changing wormhole mass.   
    
    \item As a charged solution, we would like to discuss how the electric charge of the wormhole affects the time evolution after a scalar field perturbation. As we can see in Fig.~\ref{Qchange}, with the increase of the electric charge, the high of peaks are lower while the width of peaks is broader. A wormhole with a larger electric charge could create more obvious signals of echoes.

\end{itemize}

\section{Conclusion and discussion}\label{con}

We investigated the time evolution of the symmetric and asymmetric regular black holes after a scalar field perturbation, based on three black bounce solutions. The SV metric and the LRSSV metric and the $\gamma_2=0$ case of HY metric are both describe symmetric traversable wormholes or black bounces. But if $\gamma_2\neq0$, the HY metric becomes an asymmetric solution, which causes the two sides of the wormhole throat or black bounce to be not the same.  

We considered the scalar field perturbations as a proxy for the GWs in our paper. We derived a general form of the effective potential for the spherically symmetric metric with throat-like geometric \eqref{ansazt}. We analyzed the effective potentials in both symmetric and asymmetric metrics. As a conclusion, we found there are no negative regions of the effective potentials that arise outside the black hole horizons in symmetric black bounce cases. The situations are changed in asymmetric black bounce cases, namely, they admit a negative region outside the black hole horizons. The negative regions of the effective potentials imply the instability of the systems. Thus for future work, we point out that the instabilities of asymmetric black bounce should be examined.

After getting the analytical expansions of the effective potentials of those metrics. We used the finite difference method to solve the time evolution equation of the perturbations. We numerically showed our results in Fig.~\ref{singlepeak}, \ref{gamma2change}, \ref{lchange}, \ref{Mchange} and \ref{Qchange}. As some references have pointed out that symmetric black bounce cases do not produce echo signals. Our results support this argument and showed it is also established even in the asymmetric black bounce cases. All these features characterize the properties of (a)symmetric wormholes with two peak effective potentials.

When the effective potential has two peaks, they form a well. These situations can happen in some wormhole cases. Such shapes of the effective potentials admit echoes after the perturbations. We studied the relationship of the theoretical parameter, the angular momentum of the perturbed waves, the mass, and the electric charge of the wormholes
with the properties of the echoes respectively. According to the numerical results, we may therefore conclude that massive wormholes with more charges are easier to observe echo signals after the scalar perturbations. Increases with $\gamma_2$ aggravate the asymmetry of effective potentials which leads to lower frequency but larger amplitude echoes. The angular momentum $l$ of perturbed waves changes nothing about the width of effective potentials. Larger $l$ leads to lower frequency echoes. 

As we mentioned above, the HY metric can describe RN-like black holes. We don't touch on this situation in the present paper. For future work, one can go further to investigate the time evolution after the scalar field, electromagnetic field, and gravitational field perturbations of the HY metric or other asymmetric spacetime geometries.

\section*{Acknowledgement}

We are grateful to H.L\"u, Peng Liu, Yuxuan Peng, Jingbo Yang, and De-Cheng Zou for useful discussions. This work is supported by the Initial Research Foundation of Jiangxi Normal University.

\end{document}